\documentclass[twocolumn,superscriptaddress]{revtex4-1}

\usepackage{amsbsy, amsfonts, amsmath, amssymb, amsthm, array, bm, braket ,color, dsfont, graphicx, hyperref, latexsym, layout,  mathrsfs, mathtools, siunitx, subfigure, youngtab, times, multirow}
\usepackage{bibunits}
\usepackage[dvipsnames]{xcolor}
\newcommand{\fig}[1]{Fig.~\ref{#1}}

\def\>{\rangle}
\def\<{\langle}

\newcommand{\bi}[2]{\binom{#1}{#2}}

\hypersetup{
    colorlinks=true,  
    linkcolor=blue, 
    citecolor=blue, 
    filecolor=blue,
    urlcolor=blue
}

\defaultbibliographystyle{naturemag}

\begin{document}
\title{Probabilistic one-time programs using quantum entanglement}

\author{Marie-Christine Roehsner}
\affiliation{Vienna Center for Quantum Science and Technology, Faculty of Physics, University of Vienna, Boltzmanngasse 5, 1090 Vienna, Austria}

\author{Joshua A. Kettlewell}
\affiliation{Singapore University of Technology and Design, 8 Somapah Road, Singapore 487372}
\affiliation{Centre for Quantum Technologies, National University of Singapore, 3 Science Drive 2, Singapore 117543}
\affiliation{Staple AI, 6 Raffles Quay, $\#$11-07, Singapore 048580}

\author{Joseph Fitzsimons}
\affiliation{Horizon Quantum Computing, $\#$05-22/23/24 Alice@Mediapolis, 29 Media Circle, Singapore 138565}

\author{Philip Walther}
\affiliation{Vienna Center for Quantum Science and Technology, Faculty of Physics, University of Vienna, Boltzmanngasse 5, 1090 Vienna, Austria}

\begin{abstract}

It is well known that quantum technology allows for an unprecedented level of data and software protection for quantum
computers as well as for quantum-assisted classical computers. To exploit these properties, probabilistic one-time programs have been developed, where the encoding of classical software in small quantum states enables computer
programs that can be used only once. Such self-destructing one-time programs facilitate a variety of new applications reaching from software distribution to one-time delegation of  signature authority. Whereas first proof-of-principle experiments demonstrated the feasibility of such schemes, the practical applications were limited due to the requirement of using the software on-the-fly combined with technological challenges due to the need for active optical switching and a large amount of classical communication. Here we present an improved protocol for one-time programs that resolves major drawbacks of previous schemes, by employing entangled qubit pairs. This results in four orders of magnitude higher count rates as well the ability to execute a program long after the quantum information exchange has taken place. We demonstrate our protocol over an underground fiber link between university buildings in downtown Vienna. Finally, together with our implementation of a one-time delegation of signature authority this emphasizes the compatibility of our scheme with prepare-and-measure quantum internet networks.

\end{abstract}

\maketitle

\begin{bibunit}
\section{Introduction}

Computational algorithms touch almost every aspect of modern life. In light of continuous data breaches and increasingly stricter legislation on data protection,  it would be desirable to reduce the amount of private user data leaked in a computation without forcing software owners to completely reveal their source code. Quantum computers have been shown to offer significant advantages in this area. A prominent example is blind quantum computation, where an almost classical client can delegate a quantum computation such that the quantum server cannot learn any information regarding the input, output and algorithm of the quantum computation \cite{BFK,dunjko2014,Morimae2013, Barz20012012,greganti2016}. 
 While  protocols such as this clearly demonstrate that quantum systems can provide powerful enhancements to the privacy of computations,  full scale quantum computers still present significant technical challenges.
Thus, it is of particular interest to investigate hybrid quantum-classical solutions which might allow for quantum enhancements of classical technology as well.

A promising direction of investigation is to use small quantum systems (such as single photons, which can be readily generated and manipulated by state-of-the-art quantum technology) to augment classical computers,  and in particular to use them as a resource to increase the privacy of computations. A well-known example of such hybrid systems are quantum key distribution protocols \cite{BB84, Ekert91b}.
Recently, another such hybrid system was demonstrated for probabilistic one-time programs \cite{otps}. One-time programs are a cryptographic primitive in which a server provides a client with a software in such a way that the client can obtain only one input-output pair $(x,f(x))$ before the program is destroyed. Both the input of the client and the software of the server remain private (up to the information that is leaked by the input-output pair). One-time programs are considered as a powerful building block for many cryptographic tasks and could be used for applications such as software licensing, one-time delegation of signing abilities and electronic voting schemes. It has, however, been shown that perfect information theoretically secure quantum and classical one-time programs are impossible to implement without the use of one-time self-destructing hardware (hardware which is automatically destroyed after a single use) \cite{Broadbent2013, Goldwasser2008, Liu2014}. These no-go results can be circumvented  by allowing for the possibility of error in the program outcome resulting in probabilistic one-time programs \cite{otps}.

Probabilistic one-time programs encode classical software onto single-qubit quantum states which are then sent to a client for evaluation. The client can choose the input to the program by choosing the basis they measure the  qubit in using a measurement operator taken from a set of anti-commuting operators. Thus, evaluation for one input prevents them from gaining information about a complimentary input. The output of the measurement will be the output of the gate. 
While demonstrating the implementability of probabilistic one-time programs the protocol presented in \cite{otps} faced a number of challenges with respect to theory and technological requirements that limit the practical implementation. 

One of the most important challenges in the practical implementation of any quantum communication protocol is loss tolerance. As no real-life quantum channel is without loss, in order for a protocol to be practical, it must allow for a certain level of loss of information. While the previous scheme could achieve loss tolerance, this came at the price of a sub-routine that required the program sender and receiver to implement classical back-and forth communication after the exchange of each individual qubit. Furthermore, it required the receiver of the one-time program to immediately execute the program unless they had access to a non-demolition measurement of the photon number (to perform the loss-tolerance sub-routine)  and a quantum memory. Finally, the gate rates of the scheme were limited by the mentioned need for a large amount of classical communication as well as the need of active polarization switches, resulting in a gate rate of about \si{\num{0.7}  \Hz}.

Here we present an improved protocol that exploits quantum entanglement to overcome the aforementioned limitations. We further demonstrate the enhanced practicality for real-life scenarios by using OTPs to digitally sign a message protocol using an underground fiber link that connects two buildings of the University in Vienna.
Experimentally, our protocol is based on sharing a maximally entangled qubit pair, a so-called Bell state \cite{Nielsen2000}, among the software provider (Alice) and the client (Bob), who shall use the software only once. Alice performs random measurements on her half of the Bell state that lead to the remote preparation of four different states, covering all possible one-bit software gate operations, on Bob’s side. Bob randomly chooses his measurement basis which defines input 0 or input 1. This leads to a shared table of randomly prepared input-output pairs. Now, when Alice and Bob keep their lists of respective outcomes, then Alice and Bob can use classical communication only to select the required gates for implementing the program with the corresponding input. Here it is important to point out that the execution of the program can happen long after the quantum information exchange and without any need for long-term quantum memories. Another conceptual advantage is that the use of Bell states enables the detection of a potential eavesdropper via a man-in-the-middle attack with the aim to extract information of the program. From a technological point of view this scheme allows for a strongly improved gate rate for the transmission of gates as this protocol relies only on passive optical elements. This is demonstrated by achieving gate rates of about 10 kHz after a transmission via a $650$m of  fiber link that runs partially through Vienna’s underground sewer system; corresponding to an increase in gate rate of four orders of magnitude when being compared to previous schemes that had to use active state preparation for each gate.
Besides the demonstration of delegated probabilistic one-time programs by making use of a previously established commodity table, we show that low-noise applications are possible via the implementation of one-time delegation of signature authority. We achieve success probabilities of  $>99\%$ for having Bob, as client, signing a message in  Alice’s (the program provider's) name.

\section{Theory}
\begin{figure}[b]
\centering
\includegraphics[width=0.45\textwidth]{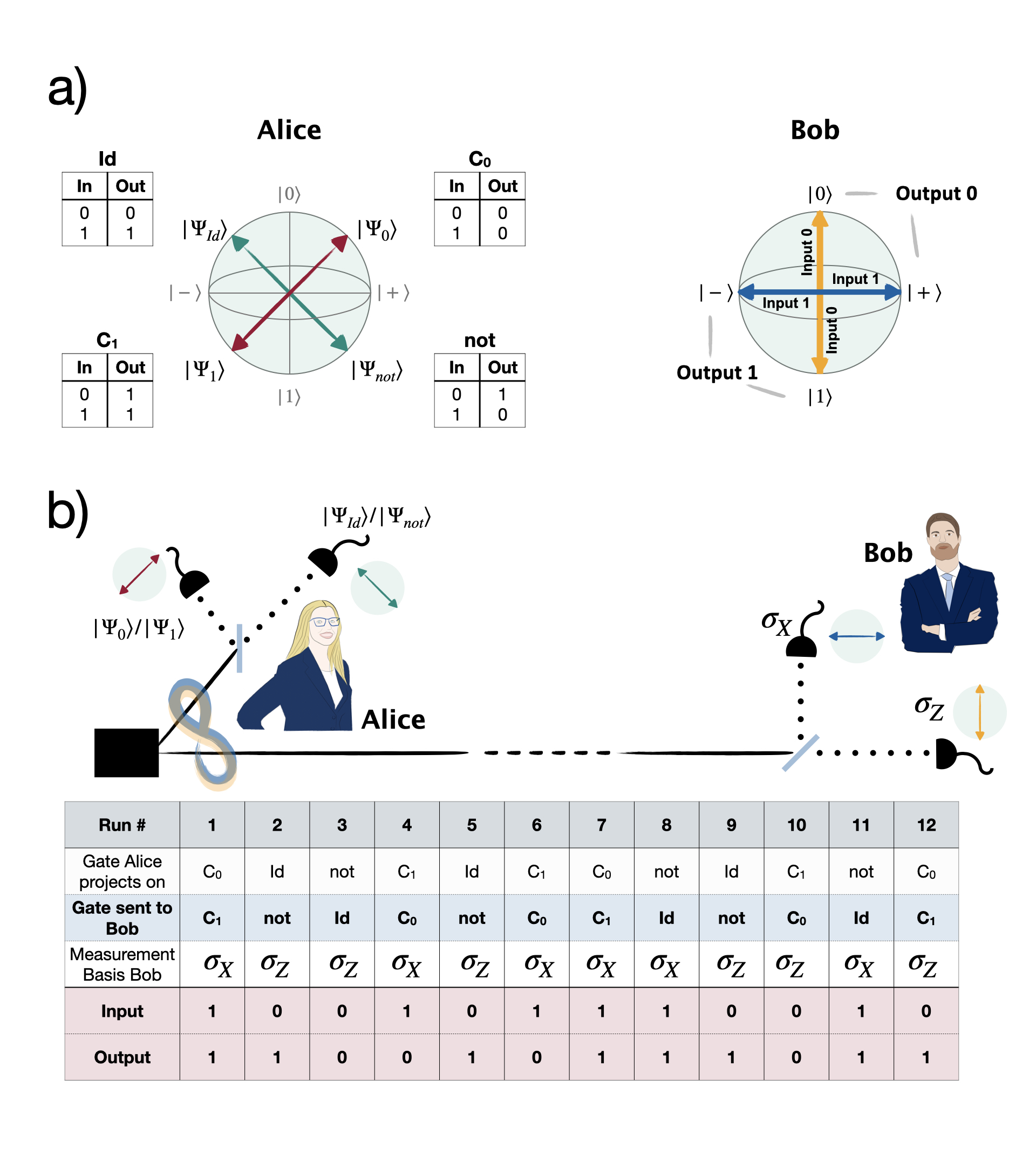}
\caption{Scheme for probabilistic one-time programs: a) shows the truth tables defining all possible 1-bit logic gates together with the quantum states representing the different $\mathcal{G}_1$ gates on the Bloch-sphere.  In b) we give the mapping of Bob's binary inputs to a measurement basis. An input of $0$ maps to a measurement in the $\sigma_Z$ (Z) basis while an input of $1$ maps to a measurement in the $\sigma_X$ (X) basis. Outputs are defined to be $0$ if Bob projects onto the positive eigenstate and $1$ for the negative eigenstate.  The success probability of the gates is given by $P_S = \frac{1}{2\sqrt{2}}+\frac{1}{2} \approx 0.85$.
b) Establishing the shared table :  Alice randomly prepares one of the four possible 1-bit gate-OTPs by randomly measuring her $\ket{\Psi^-}$ Bell-state in one of the two bases given by $\ket{\Psi_0}/\ket{\Psi_1}$ and $\ket{\Psi_{Id}}/\ket{\Psi_{not}}$. This collapses Bob's qubit into the orthogonal state which is sent over a quantum channel to Bob. He will randomly measure in $\sigma_Z$ or $\sigma_X$, corresponding to a random input of $0$ or $1$ to the gate. Alice notes the gates sent (blue shaded row) and Bob the inputs and outputs of the gate (red shaded row). These (classical) records form the shared table which will later be used to execute a program. Alice and Bob repeat this procedure until a sufficient amount of gate-OTPs has been exchanged. To increase clarity of the illustrations the gates are shown here with a $100\%$ success probability. In a real implementation Bob will receive the correct output with a probability of $P_S \approx 0.85$.} 
\label{scheme}
\end{figure}

We define a one-time program as follows: a sender or provider, Alice,  supplies resources related to a function $f(\cdot)$ to a receiver or client, Bob, which allow him to evaluate $f(x)$ while gaining no knowledge of $f(x')$, for any $x'\neq x$, other than what is directly implied by $f(x)$. Alice in turn obtains no information regarding Bob's input $x$. As it was shown that perfect, information theoretically secure OTPs in the classical and quantum case are impossible without further assumptions on hardware or abilities of an adversary  \cite{1997insecurity, Goldwasser2008, Broadbent2013} we allow for a bounded probability of error in the program output when encoding classical software onto quantum states \cite{otps} yielding probabilistic one-time programs. To achieve these probabilistic OTPs we start with the most basic logical gates which map a single input bit to a single output bit. We will refer to these gates as $\mathcal{G}_1$ gates. Alice will encode her choice of gate onto a quantum state (typically a qubit) while Bob's input will correspond to his measurement basis.
We choose to encode Bob's input as measurement in $\sigma_Z$ for an input of $0$ and a measurement in $\sigma_X$ for an input of $1$. This allows Alice to, probabilistically, encode the four possible $\mathcal{G}_1$ gates (Identity, NOT, Constant-Zero and Constant-One) as one of the following four single-qubit quantum states (truth tables of the gates and Bloch-sphere representation of the states are shown in \fig{scheme}a): 
%
\begin{subequations}
\begin{align}
\ket{\Psi_0} = \frac{1}{\sqrt{2+\sqrt{2}}} (\ket{0} + \ket{+} )\; \label{linear1}\\
\ket{\Psi_1} = \frac{1}{\sqrt{2+\sqrt{2}}} (\ket{1} - \ket{-}) \; \label{linear2} \\
\ket{\Psi_{{Id}}} = \frac{1}{\sqrt{2+\sqrt{2}}} (\ket{0} + \ket{-} )\;  \label{linear3} \\
\ket{\Psi_{{not}}} = \frac{}{\sqrt{2+\sqrt{2}}} (\ket{1} + \ket{+} ) \; 
\label{linear4}
\end{align}
\end{subequations}
%
where $\ket{\pm} = \frac{1}{\sqrt{2}}\left( \ket{0} \pm \ket{1} \right)$. 
 
It has been shown \cite{otps} that these four basic gates allow for universal classical computing when being combined with a larger classical circuit in a fixed configuration. Remarkably, the circuit arrangements can be public as only the basic gate operations need to be secret to hide the implemented software.

In the original OTP scheme the basic gates were consecutively mapped onto quantum states, realized as single photons, and then send to a receiver that had to measure them in the same exact order for implementing the function (software). The scheme presented here exploits quantum entanglement to create randomness as a resource for an enhanced protocol that allows to share OTPs that can be used at any time.

The new protocol is composed of two distinct parts: a quantum part in which Alice sends a random sequence of gate-OTPs which Bob measures in a random basis and a classical part where classical communication is used to implement a OTP using the previously shared information from the quantum part.
To randomly prepare one of the gate states ($\ket{\Psi_0}, \ket{\Psi_1}, \ket{\Psi_{Id}}, \ket{\Psi_{not}}$) Alice generates a maximally entangled Bell-state, measures it randomly in one of two bases ($\ket{\Psi_0}/\ket{\Psi_1}$) or ($\ket{\Psi_{Id}}/\ket{\Psi_{not}}$) which  leads to a remote state generation on the other qubit that is sent to Bob. Thus, Bob will receive a random gate-OTP which he will randomly measure in $\sigma_Z$ (input $0$) or $\sigma_X$ (input $1$). Alice notes the gates sent while Bob records the input-output pairs, and both keep their results private. 
The remaining events make up a table of imperfectly correlated results where the percentage of correct input-output pairs is given by $P_1 = \frac{1}{2\sqrt{2}}+\frac{1}{2} \approx 0.85$. This is referred to as a \textit{shared table} and will later be used as classical commodity or resource to perform a program.  Alice and Bob repeat this process until they have constructed a shared table of sufficient length for the program(s) they want to perform. Note that this results in the sequence of gates being independent and identically distributed (IID).
In case of channel losses Alice and Bob will exchange information about when they have sent and received qubits, repectively. Only the results of events in that both parties agree that a qubit was sent and measured will be kept, all other results will be discarded. Thus, channel losses do not affect the security of the protocol, as only coincidence events generate entries in the shared table and other events will not be used for the protocol.

\begin{figure}[ht]
\centering
\includegraphics[width=0.45\textwidth]{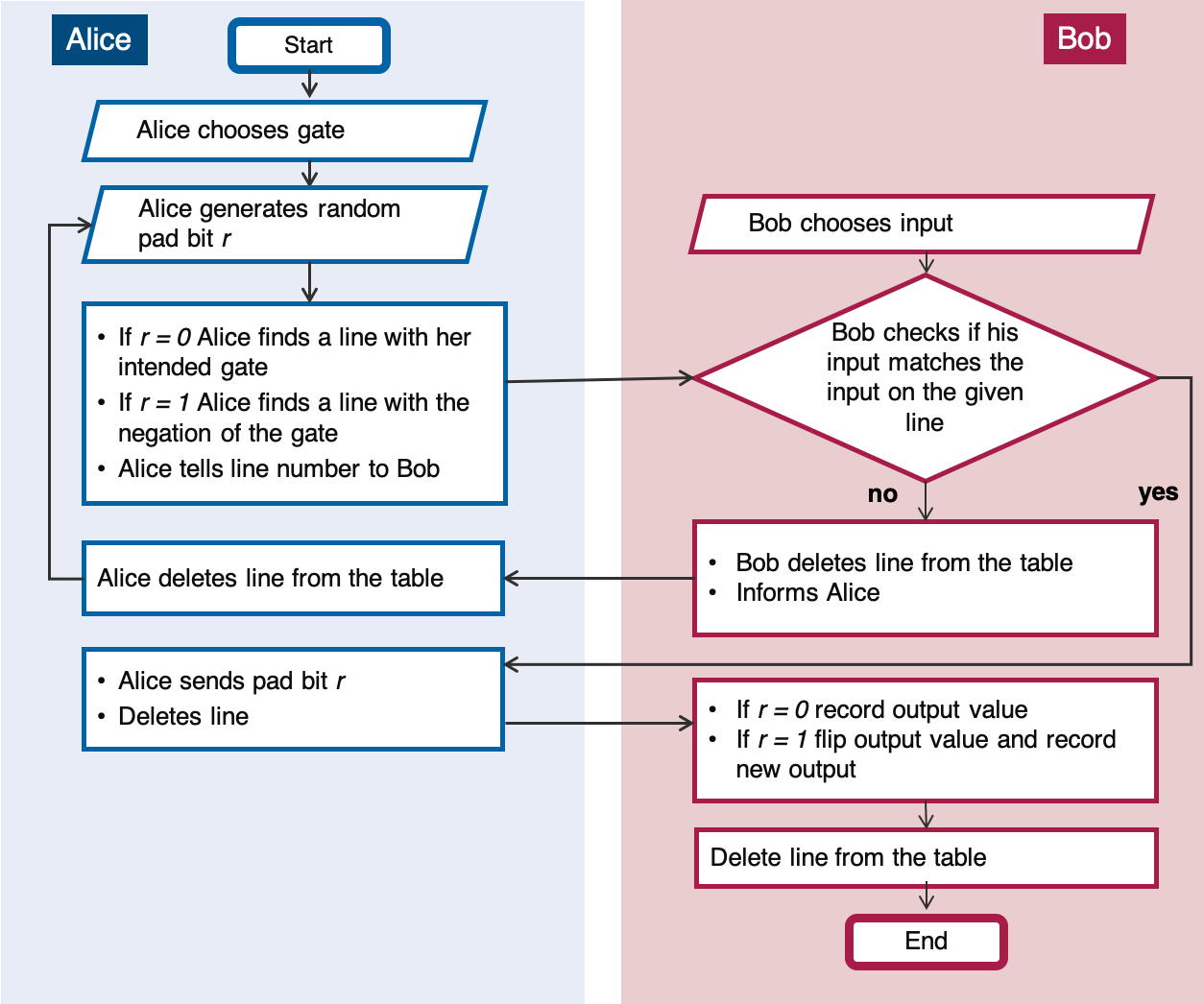}
\caption{Flow chat showing the instructions for Alice and Bob to securely evaluate a single $\mathcal{G}_1$ gate-OTP. The runtime of the classical part scales linearly with the complexity of function and latency between the parties as shown in the appendix. } 
\label{flow}
\end{figure}

After the distribution the shared table can be used to run an OTP (see flow chart in \fig{flow}). To execute a gate Alice will first generate a random bit $r$. If $r=0$ she looks at her part of the shared table and finds a line with the desired gate, if $r=1$ she finds a line with the opposite gate (i.e. the gate for which all outputs are flipped). Lines she skips over while looking for an appropriate gate will be deleted from the table. She will then ask Bob if he can use this line. If Bob's desired input is equal to the (random) input in that line of the table, he will accept. Otherwise he will decline the use of the line and they will repeat the process (using a newly generated $r$). Only when Bob accepts to use a line Alice will reveal the corresponding value of $r$. If $r=1$ Bob will have to flip the result of the gate used. Once a line is used (accepted or declined) it will be deleted from the shared table. Alice and Bob will iterate this process until the desired circuit is completed.  The use of the random one-time-pad ($r$) to encrypt Bob's output prevents information leakage in the case that Bob chooses to not use any given line (honestly or dishonestly) as without the line's pad value he gains no information. If the individual gates are used as building blocks for a larger circuit Alice might be concerned about Bob learning the intermediate results of this circuit.  She can prevent this by randomly inserting pairs of NOT gates, with a probability of $1/2$, between the gates and subsequently absorbing them into the neighbouring gates as described in \cite{otps}. This will not alter the outcome of the overall program but effectively apply a one-time pad on the intermediate results of the circuit. 

Remarkably, the quantum channel connecting Alice and Bob only needs to be maintained for the period required to generate the shared table. Thus, the creation of the shared table may occur long before the classical communication to execute a program and a large shared table might be used to execute several programs.

Furthermore, we can consider the situation where an eavesdropper might perform a man-in-the-middle attack to steal the OTP. Such an eavesdropper attempting to intercept the program would need to be present in both, the quantum and the classical channel to recreate all steps of the protocol. Intercepting the classical channel alone is of no use as the implementation of a gate OTP requires also the knowledge of the shared table. However, for obtaining this knowledge an eavesdropper must intercept the quantum channel by measuring and resending the quantum states. In analogy to entanglement-based quantum cryptography protocols \cite{Ekert91b} this can be detected by Alice and Bob when using a subset of their shared table’s lines for evaluating a Clauser-Horne-Shimony-Holt (CHSH) Bell inequality \cite{Clauser1969}.
Advantageously, the measurement settings for obtaining the CHSH-Bell-parameter can be directly extracted from the used setting for implementing the OTP. Thus Alice and Bob just need to choose at the time when the program is executed which rows of the shared table  should be taken for detecting a potential eavesdropper.

Finally, we would like to note that our OTPs are equivalent to noisy examples of random $\bi{2}{1}$-oblivious transfer (OT), a versatile cryptographic resource allowing a user to access a subset of database entries or messages a sender transmits without the sender knowing which entry was accessed. OT is known to be sufficient for many secure multi-party processes \cite{beaver1, Kilian, Yuval} such as homomorphic encryption \cite{SemihomomorphicEncryption} and bit commitment \cite{Yao86}.  Classically, OT may only be performed with assumptions on the computational power of the parties \cite{Impagliazzo89} and is known to be impossible to implement even with quantum computers when information theoretic security is required \cite{1997insecurity}.


\section{Experimental Implementation}
We experimentally demonstrated our entanglement-based one-time programs between two university buildings separated by approximately \si{\num{200}\meter} air-line distance (approximately \si{\num{650}\meter} in fiber) in down-town Vienna.

\begin{figure}[h]
\centering
\includegraphics[width=0.5\textwidth]{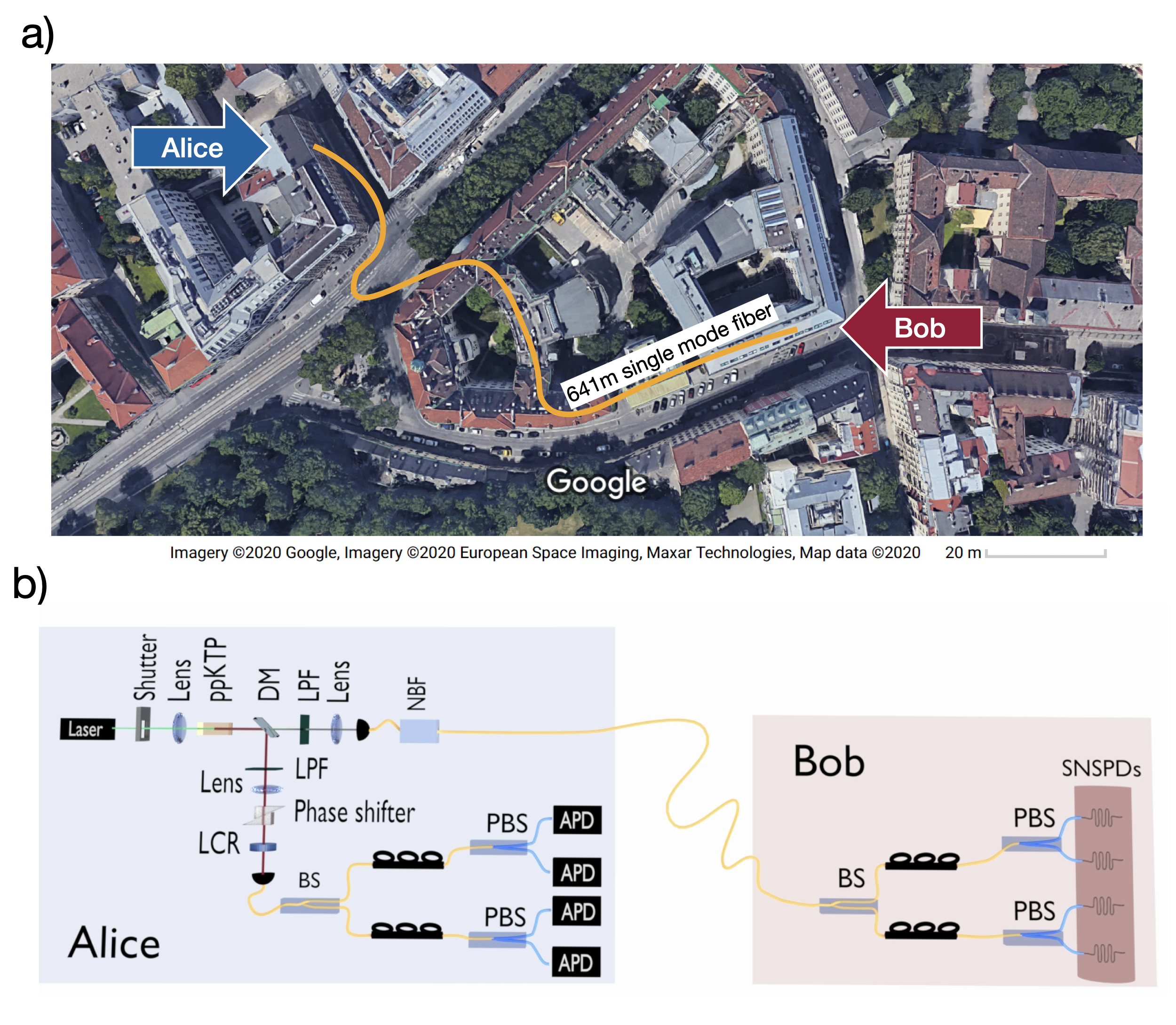}
\caption{Experimental setup and approximate fiber path connecting Alice and Bob: a) The laboratories of Alice and Bob are located at different buildings of the University in Vienna, but connected by a quantum channel consisting of a single-mode fiber (Corning SMF-28) with a lenght of 641m.  b) Polarization entangled photons are created via SPDC using a 515nm cw-pump-laser directed on to a ppKTP crystal, emitting polarisation-entangled photon pairs in a $\ket{\Psi^-}=\frac{1}{\sqrt{2}}(\ket{H}_s\ket{V}_l+e^{i\theta}\ket{V}_s\ket{H}_l)$ Bell state with $\lambda_s=785nm$ and $\lambda_l=1498nm$ (Telecom S-Band). A dichroic mirror (DM) separates the photons by wavelength followed by a long-pass filter (LPF) to block the pump light in both arms and a narrow bandwidth filter (NBF), to ensure spectral indistinguishability of the $H$ and $V$ photons, in the long-wavelength arm. In the short-wavelength arm we use calcite wedges as phase shifters to compensate temporal walk-off and a liquid crystal retarder (LCR) to set precisely the phase $\theta$ in the Bell state. Alice uses a beam-splitter (BS) to randomly choose her measurement basis. The measurement is realized using in-fiber polarization control, two fiber-polarizing beam-splitters (PBS) and four Si-Avalanche-Photo-Diodes (APD). The second photon is transmitted through $\approx 650 m$ of standard Telecom fibre to Bob's laboratory. Bob uses one fiber-BS, in-fiber polarization control, two PBSs and four superconducting nanowire single-photon detectors (SNSPDs) to randomly measure the received photons in one of two bases $\sigma_Z$ and $\sigma_X$ corresponding to input $0$ and $1$ respectively.}                                                                                                                                                                                                                                      
\label{setup}
\end{figure}

Alice initially prepares a maximally entangled Bell-state $\ket{\Psi^-}=\frac{1}{\sqrt{2}}(\ket{0}\ket{1}-\ket{1}\ket{0})$. She keeps one of the qubits in her local laboratory and sends it to a $50/50$-beamsplitter. One output arm of the beamsplitter leads to a measurement device configured to measure in the basis spanned by $\ket{\Psi_0}$ and $\ket{\Psi_1}$, while qubits leaving the beamsplitter in the other output will be measured in the basis given by $\ket{\Psi_{Id}}$ and $\ket{\Psi_{not}}$. %
The second qubit is sent to Bob through a quantum channel, which is realized by a standard telecom fiber that is located partially in Vienna's sewer system. Bob uses a similar measurement apparatus as Alice, which also relies on a $50/50$-beamsplitter leading to measurement devices projecting in $\sigma_Z$ and $\sigma_X$ bases. Both Alice and Bob record their measurement results and thus the gates send respectively the input and output of the program, which allows them to generate the shared table.  A scheme of our set-up is shown in \fig{setup}.

To prepare the Bell state Alice uses a novel photon source design (adapted from \cite{Laudenbach2017}) for generating entangled photon pairs with tailored wavelengths such that the transmitted photon faces minimal absorption loss in fiber and that the local photon can be efficiently detected with standard detector technology. This  single-pass spontaneous parametric down conversion (SPDC) source   emits highly non-degenerate polarization entangled photons pairs in the $\ket{\Psi^-}=\frac{1}{\sqrt{2}}(\ket{H}_s\ket{V}_l-\ket{V}_s\ket{H}_l)$ Bell state, where $\ket{H}$ corresponds to horizontal and $\ket{V}$ to vertical polarisation and $s$ and $l$ denote the short ($785$nm) and long ($1498$nm) wavelength path. In our source two down-conversion processes are phase-matched in the same crystal yielding photon pairs of  $\ket{H}_s\ket{V}_l$ as well as  $\ket{V}_s\ket{H}_l$ polarization. These are superimposed to create a Bell-state of the form $\frac{1}{\sqrt{2}}(\ket{H}_s\ket{V}_l+e^{i\theta}\ket{V}_s\ket{H}_l)$. To ensure the spectral indistinguishably of these two processes a tunable narrow bandwidth-filter is inserted in the long wavelength path. A phase shifter is used to compensate for the varying time delays due to mismatched group velocities in the crystal and a liquid crystal retarder is used to set the phase angle $\theta$ of the produced Bell state. The pump-wavelength and crystal poling-period were chosen such that the source emits one photon in the Telecom range at $1498$nm (Telecom S-Band) and the other photon in the near-infrared range at $785$nm which is a standard wavelength for optical manipulation and in particular for efficient detection by using commercial Silicon Avalance Photo-Diodes (APDs). The Telecom ($1498$nm) photon is sent through approximately $650$m of fiber to Bob's laboratory where they are detected by superconducting nano-wire detectors, as this wavelength suffers from low losses in fiber transmission. This results in a coincidence and thus gate rate of $10$kHz corresponding to an improvement in gate rate by four orders of magnitude compared to the previous implementation \cite{otps}. 

%
%
%
%

\section{Implemented Program}
\begin{figure}[h]
\centering
\includegraphics[width=0.48\textwidth]{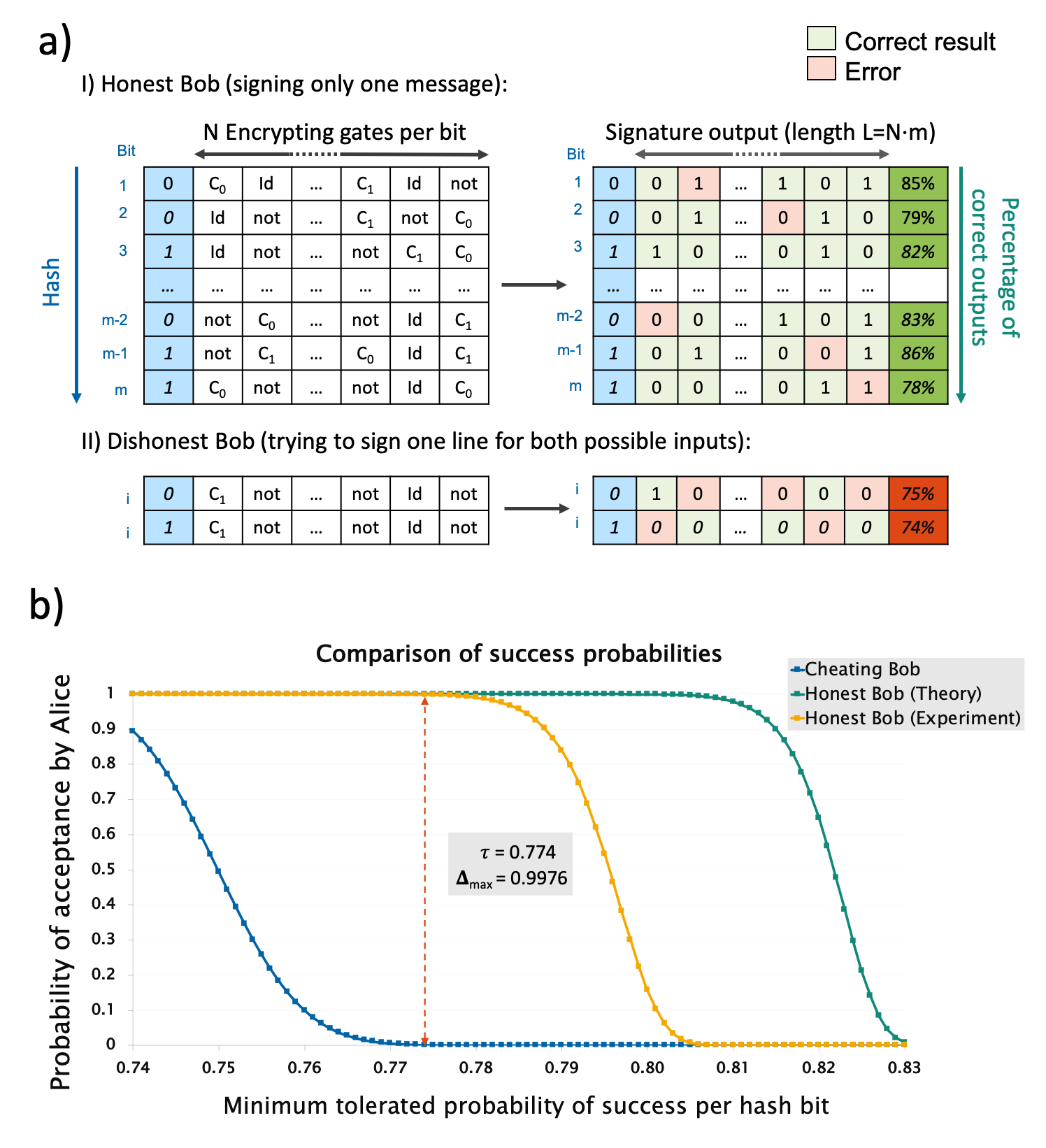}
\caption{Delegated signature protocol and comparison of the success probabilities for different scenarios. a)  Evaluation of the signature. Alice and Bob evaluate $L = N \cdot m$ lines from the shared table, according to the hash of Bobs messages. The signature produces a $N$ bit string for each bit of the hash, each one required to be correct in $\tau \cdot N$ positions, where the correct output is defined as a ideal implementation of the gate. $\tau$ is chosen according to the length of $N$ to maximise the difference between honest and dishonest probabilities. If Bob tries to cheat and sign two messages that differ in only one bit of the hash he has to obtain two sets of correct outputs for N gates (corresponding to one line). This will reduce his average success probability as shown in \cite{otps} and thus his probability to surpass Alice's threshold.  b) The probabilities of signing 1 (honest) or 2 (dishonest) messages using a signature length of $N=1000$ and a hash output size of 224 bits. The difference between the honest (experimental) and dishonest probability of success is maximised (at $0.9976$) for the experimentally found values for a threshold value of $\si{\num{77.4}\percent}$ (indicated by the red dotted line) which corresponds to a success probability of $0.9987$ and a cheating probability of $0.0011$. }                                                                                                                                                                                                                                      
\label{Signature}
\end{figure}

We show the experimental implementation of a protocol for one-time delegation of signature authority  in which Alice enables Bob to sign exactly one message in her name. While in general the complexity of programs that can be implemented by our approach is limited by their probabilistic nature, this protocol's success probability can be increased (in principle arbitrarily close to 1) without compromising the security \cite{otps}. Digital signatures are a widely employed technique used for contract signing, software distribution, e-mails and  numerous other applications. Sometimes it is desirable to delegate these capabilities (e.g. to a lawyer), which classically corresponds to handing over one's private key. However, this enables the recipient to sign an unlimited number of messages as the classical software used for signing can, in principle, always be copied. Thus, should one wish to limit the number of messages that can be signed, this cannot be done classically. OTPs on the other hand enable us to implement a one-time delegated signatures as introduced in \cite{otps} with information theoretic security following the described steps:

\begin{enumerate}
\item Encryption: Alice prepares a set of OTPs that will perform encryption with her private key(s). As the encryption will be done bitwise it is sufficient to use $\mathcal{G}_1$ gate-OTPs in this step. She sends these gate-OTPs over to Bob. For every bit that Bob wants to encrypt, Alice will send $N$ independently encrypting gate-OTPs. These will result in multiple independent encryptions which will later allow her to achieve an increased probability of success.
\item Message: Bob chooses the message he wants to sign in Alice's name. As in classical digital signatures he takes the hash of this message which ensures that his input into the protocol will always be of the same length $m$. 
\item Signing: Bob uses the bits of his hash as inputs into the gate-OTPs. The output of the gate-OTPs will form the delegated signature. As he receives $N$ gate-OTPs per bit of the hash, the length of the signature will be $L = m \cdot N$.
\item Verification: Bob sends the signature together with the signed message back to Alice for verification. Alice will accept the signature as valid if the expected percentage of output bits is correct. Thus, she will define a lower bound or threshold $\tau$ on the probability of success she will accept for the encryption of every individual bit in the hash. Should one or more bits of the hash have been signed with a probability of success below her threshold she will abort the protocol (see also \fig{Signature}b).

\end{enumerate}
%

Intrinsically the individual gate-OTPs have a success probability of $P_S = \frac{1}{2\sqrt{2}}+\frac{1}{2} \approx 0.85$.
The overall success probability of the protocol  is however increased by using multiple gate-OTPs per bit of the hash.  In fact, by increasing $N$ the probability that at least $\tau \cdot N$ evaluations are correct (i.e. the success probability of the signature) asymptotically approaches $1$. It is important to note that in order to maintain the security of the protocol the $N$ gate-OTPs that are used per bit of the hash are not mere copies of each other but encrypt the bit independently, i.e. with a different private key.

\begin{figure}[h]
\centering
\includegraphics[width=0.48\textwidth]{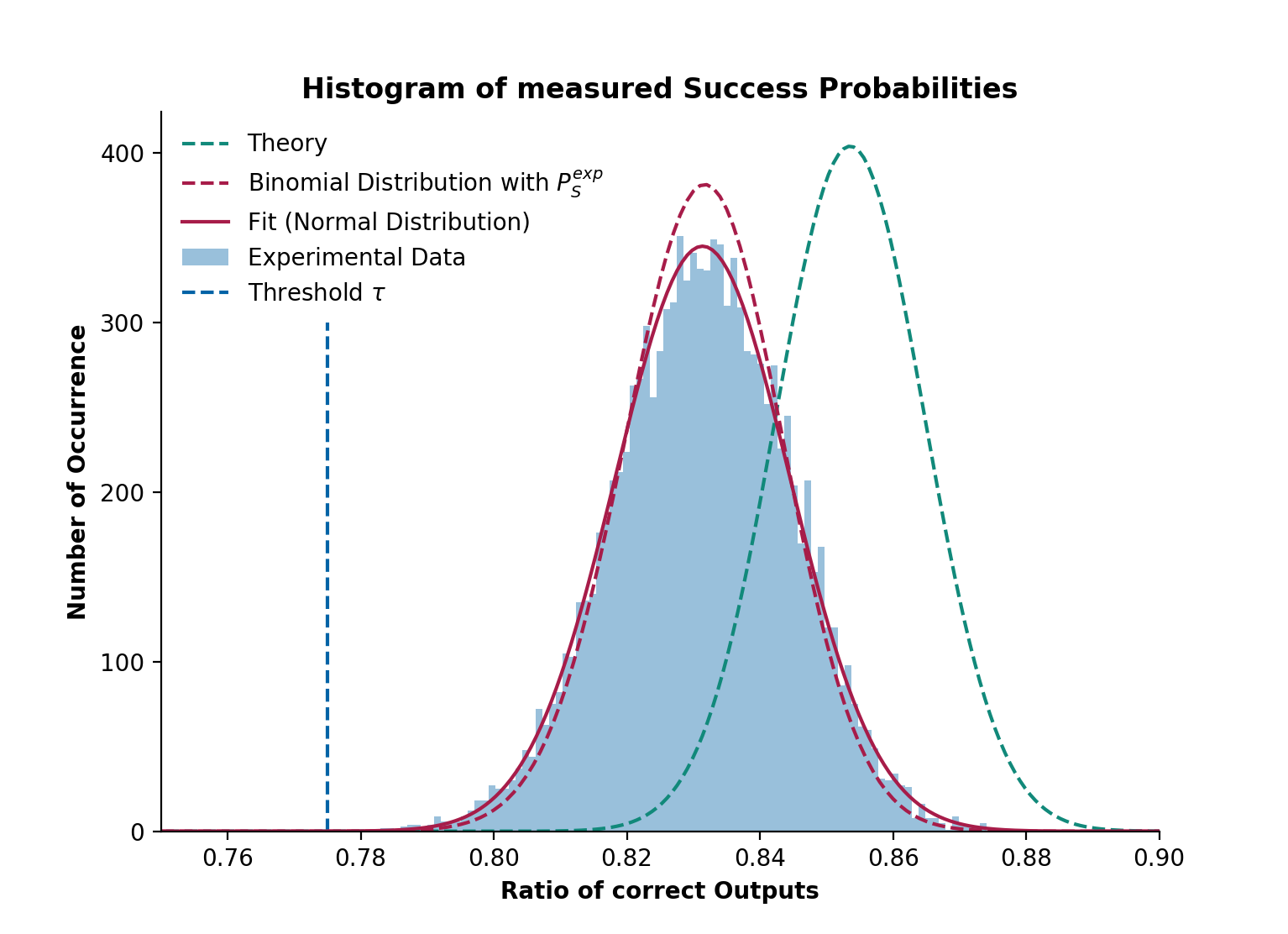}
\caption{Cumulative histogram of success probabilities in $50$ experimental implementations of delegated signatures. The light blue bars show the experimentally found probabilities of success per bit of the hash for $50$ signatures with $224$ hash-bits each, thus from $11 200$ evaluations (with $N=1000$, thus $112,000,000$ evaluated gate-OTPs). While due to experimental imperfections the probability of success is lower than the theoretical maximum, nevertheless Alice's acceptance threshold $\tau$ is passed every single time. To characterize the found distribution we compare it the theoretical (noiseless) prediction (dashed green line) as well as to a binomial distribution with the experimentally found mean $\mu^{exp} = 0.831 $ (red dashed line) and a fit to the histogram (normal distribution, red solid line, $\mu^{exp} = 0.831, \sigma_{exp}=0.013 $).  We attribute the slightly increased standard deviation in the data compared to the binomial distribution to drifts in the set-up during data acquisition. }                                                                                                                                                                                                                                      
\label{Histogram}
\end{figure}

Experimentally we implemented a one-time delegated signature using $N= 1000$ and a SHA3-224 hash ($m=224$), thus per signature we evaluate $L = N \cdot m= 224 000$ gate-OTPs. Due to experimental imperfections the  probability of success is reduced to $P_s^{exp} = 0.831 \pm 0.013$ . Given these values we analysed the probability of success for a honest Bob, trying so sign one-and-only-one message, compared to a cheating Bob. To bound the probability of successful cheating we assume the smallest deviation and thus worst-case in which Bob tries to sign a second  message that differs in just one bit of the hash from his first message. We consider his probability of success in dependence of Alice's threshold value $\tau$. Furthermore, we assume that a cheating Bob can achieve the theoretical maximum for $P_S^{th}=\frac{1}{2\sqrt{2}}+\frac{1}{2} \approx 0.85$. Thus, unless Bob can exploit a collision in the classical hash, our values give an upper bound for his probability of success.
Considering these numbers we choose  $\tau = 0.776$ to maximise the difference in probability of success between an honest and a cheating Bob as shown in \fig{Signature} where we plot the respective success probabilities in dependence of $\tau$.
At this value a cheating Bob has a probability of success of $P_{cheat} = 0.11\%$ while an honest Bob achieves $P_{hon} = 99.87\%$.
In figure \fig{Histogram} we show a histogram of the combined results of 50 (honest) delegated signatures (corresponding to $11,200,000$ evaluated gate OTPs) where each bar is generated using the results of $1000$ OTPs. It can be seen that due to experimental imperfections and drifts in the setup that the average success probability is lower than the theoretical maximum (green line) and has a larger standard deviation than expected by a binomial distribution of this mean (red line). Nevertheless, the protocol is successfully implemented and the threshold of acceptance by Alice is surpassed every single time.

The evaluation of $L$ gate-OTPs would trivially require evaluating $L$ rounds of communication to complete. However, as none of the gates in the signature scheme are causally connected, Alice and Bob may evaluate all of them concurrently, thus reducing the expected required rounds of communication to $\log_2 (L)$ where $L$ is the total number of gate-OTPs. Therefore, on average our example program could be implemented using only $18$ rounds of classical communication.  Should Alice and Bob be willing to use  $O(\log_2(L))$ lines per input the amount of communication rounds can be made constant with a high probability.

\section{Conclusion}
We have presented a new protocol for probabilistic one-time programs overcoming previous challenges in theory and experiment. Our implementation exploits quantum entanglement as a resource to achieve random remote state preparation resulting in a shared table of correlated input-output pairs between Alice and Bob. Through separating the quantum communication from the actual program execution, we enable client and sender to perform a one-time program at an arbitrarily later time only using classical communication. 
By deploying our experiment between two university buildings, connected by an underground quantum link we demonstrate the significant advantages of this method over the previous state of the art, allowing for four orders of magnitude higher gate-rates than in previous experiments. Additionally, the use of quantum entanglement enables the detection of an eavesdropper, attempting to steal the program. We believe that can be the basis for a wide field of further investigations including new protocols and connections to known protocols like oblivious transfer and quantum money \cite{guan2018,bozzio2018, erven2014}. Further advances in source and detector technologies, would allow gate rates to be increased even further. We believe that this demonstration indicates the compatibility of our schemes with early quantum internet implementations and highlights the viability of quantum technologies using small quantum systems to enhance our current classical capabilities. 


\begin{thebibliography}{10}
\expandafter\ifx\csname url\endcsname\relax
  \def\url#1{\texttt{#1}}\fi
\expandafter\ifx\csname urlprefix\endcsname\relax\def\urlprefix{URL }\fi
\providecommand{\bibinfo}[2]{#2}
\providecommand{\eprint}[2][]{\url{#2}}

\bibitem{BFK}
\bibinfo{author}{Broadbent, A.}, \bibinfo{author}{Fitzsimons, J.} \&
  \bibinfo{author}{Kashefi, E.}
\newblock \bibinfo{title}{Universal blind quantum computation}.
\newblock In \emph{\bibinfo{booktitle}{Proceedings of the 50th Annual Symposium
  on Foundations of Computer Science}}, \bibinfo{pages}{517--526}
  (\bibinfo{year}{2009}).

\bibitem{dunjko2014}
\bibinfo{author}{Dunjko, V.}, \bibinfo{author}{Fitzsimons, J.~F.},
  \bibinfo{author}{Portmann, C.} \& \bibinfo{author}{Renner, R.}
\newblock \bibinfo{title}{Composable security of delegated quantum
  computation}.
\newblock In \emph{\bibinfo{booktitle}{International Conference on the Theory
  and Application of Cryptology and Information Security}},
  \bibinfo{pages}{406--425} (\bibinfo{organization}{Springer},
  \bibinfo{year}{2014}).

\bibitem{Morimae2013}
\bibinfo{author}{Morimae, T.} \& \bibinfo{author}{Fujii, K.}
\newblock \bibinfo{title}{Blind quantum computation protocol in which alice
  only makes measurements}.
\newblock \emph{\bibinfo{journal}{Phys. Rev. A}} \textbf{\bibinfo{volume}{87}},
  \bibinfo{pages}{050301} (\bibinfo{year}{2013}).

\bibitem{Barz20012012}
\bibinfo{author}{Barz, S.} \emph{et~al.}
\newblock \bibinfo{title}{Demonstration of blind quantum computing}.
\newblock \emph{\bibinfo{journal}{Science}} \textbf{\bibinfo{volume}{335}},
  \bibinfo{pages}{303--308} (\bibinfo{year}{2012}).

\bibitem{greganti2016}
\bibinfo{author}{Greganti, C.}, \bibinfo{author}{Roehsner, M.-C.},
  \bibinfo{author}{Barz, S.}, \bibinfo{author}{Morimae, T.} \&
  \bibinfo{author}{Walther, P.}
\newblock \bibinfo{title}{Demonstration of measurement-only blind quantum
  computing}.
\newblock \emph{\bibinfo{journal}{New Journal of Physics}}
  \textbf{\bibinfo{volume}{18}}, \bibinfo{pages}{013020}
  (\bibinfo{year}{2016}).

\bibitem{BB84}
\bibinfo{author}{Bennett, C.~H.} \& \bibinfo{author}{Brassard, G.}
\newblock \bibinfo{title}{Quantum cryptography: Public key distribution and
  coin tossing}.
\newblock In \emph{\bibinfo{booktitle}{Proceedings of IEEE International
  Conference on Computers, Systems and signal processing}}, vol.
  \bibinfo{volume}{175}, \bibinfo{pages}{8} (\bibinfo{year}{1984}).

\bibitem{Ekert91b}
\bibinfo{author}{Ekert, A.~K.}
\newblock \bibinfo{title}{Quantum cryptography based on bell's theorem}.
\newblock \emph{\bibinfo{journal}{Phys. Rev. Lett.}}
  \textbf{\bibinfo{volume}{67}}, \bibinfo{pages}{661--663}
  (\bibinfo{year}{1991}).

\bibitem{otps}
\bibinfo{author}{Roehsner, M.-C.}, \bibinfo{author}{Kettlewell, J.~A.},
  \bibinfo{author}{Batalh{\~a}o, T.~B.}, \bibinfo{author}{Fitzsimons, J.~F.} \&
  \bibinfo{author}{Walther, P.}
\newblock \bibinfo{title}{Quantum advantage for probabilistic one-time
  programs}.
\newblock \emph{\bibinfo{journal}{Nature communications}}
  \textbf{\bibinfo{volume}{9}}, \bibinfo{pages}{1--8} (\bibinfo{year}{2018}).

\bibitem{Broadbent2013}
\bibinfo{author}{Broadbent, A.}, \bibinfo{author}{Gutoski, G.} \&
  \bibinfo{author}{Stebila, D.}
\newblock \bibinfo{title}{Quantum one-time programs}.
\newblock In \bibinfo{editor}{Canetti, R.} \& \bibinfo{editor}{Garay, J.~A.}
  (eds.) \emph{\bibinfo{booktitle}{Advances in Cryptology -- CRYPTO 2013: 33rd
  Annual Cryptology Conference, Santa Barbara, CA, USA, August 18-22, 2013.
  Proceedings, Part II}}, \bibinfo{pages}{344--360}
  (\bibinfo{publisher}{Springer Berlin Heidelberg}, \bibinfo{address}{Berlin,
  Heidelberg}, \bibinfo{year}{2013}).

\bibitem{Goldwasser2008}
\bibinfo{author}{Goldwasser, S.}, \bibinfo{author}{Kalai, Y.~T.} \&
  \bibinfo{author}{Rothblum, G.~N.}
\newblock \emph{\bibinfo{title}{One-Time Programs}}, \bibinfo{pages}{39--56}
  (\bibinfo{publisher}{Springer Berlin Heidelberg}, \bibinfo{address}{Berlin,
  Heidelberg}, \bibinfo{year}{2008}).

\bibitem{Liu2014}
\bibinfo{author}{Liu, Y.-K.}
\newblock \bibinfo{title}{Single-shot security for one-time memories in the
  isolated qubits model}.
\newblock In \bibinfo{editor}{Garay, J.~A.} \& \bibinfo{editor}{Gennaro, R.}
  (eds.) \emph{\bibinfo{booktitle}{Advances in Cryptology -- CRYPTO 2014}},
  \bibinfo{pages}{19--36} (\bibinfo{publisher}{Springer Berlin Heidelberg},
  \bibinfo{address}{Berlin, Heidelberg}, \bibinfo{year}{2014}).

\bibitem{Nielsen2000}
\bibinfo{author}{Nielsen, M.~A.} \& \bibinfo{author}{Chuang, I.~L.}
\newblock \emph{\bibinfo{title}{Quantum Computation and Quantum Information}}
  (\bibinfo{publisher}{Cambridge University Press},
  \bibinfo{address}{Cambridge}, \bibinfo{year}{2000}).

\bibitem{1997insecurity}
\bibinfo{author}{{Lo}, H.-K.}
\newblock \bibinfo{title}{{Insecurity of quantum secure computations}}.
\newblock \emph{\bibinfo{journal}{Physical Review a.}}
  \textbf{\bibinfo{volume}{56}}, \bibinfo{pages}{1154--1162}
  (\bibinfo{year}{1997}).
\newblock \eprint{quant-ph/9611031}.

\bibitem{Clauser1969}
\bibinfo{author}{Clauser, J.~F.}, \bibinfo{author}{Horne, M.~A.},
  \bibinfo{author}{Shimony, A.} \& \bibinfo{author}{Holt, R.~A.}
\newblock \bibinfo{title}{Proposed experiment to test local hidden-variable
  theories}.
\newblock \emph{\bibinfo{journal}{Phys. Rev. Lett.}}
  \textbf{\bibinfo{volume}{23}}, \bibinfo{pages}{880--884}
  (\bibinfo{year}{1969}).

\bibitem{beaver1}
\bibinfo{author}{Beaver, D.}
\newblock \bibinfo{title}{Precomputing oblivious transfer}.
\newblock In \bibinfo{editor}{Coppersmith, D.} (ed.)
  \emph{\bibinfo{booktitle}{Advances in Cryptology --- CRYPT0' 95}},
  \bibinfo{pages}{97--109} (\bibinfo{publisher}{Springer Berlin Heidelberg},
  \bibinfo{address}{Berlin, Heidelberg}, \bibinfo{year}{1995}).

\bibitem{Kilian}
\bibinfo{author}{Kilian, J.}
\newblock \bibinfo{title}{Founding crytpography on oblivious transfer}.
\newblock In \emph{\bibinfo{booktitle}{Proceedings of the Twentieth Annual ACM
  Symposium on Theory of Computing}}, STOC '88, \bibinfo{pages}{20--31}
  (\bibinfo{publisher}{ACM}, \bibinfo{address}{New York, NY, USA},
  \bibinfo{year}{1988}).

\bibitem{Yuval}
\bibinfo{author}{Ishai, Y.}, \bibinfo{author}{Prabhakaran, M.} \&
  \bibinfo{author}{Sahai, A.}
\newblock \bibinfo{title}{Founding cryptography on oblivious transfer --
  efficiently}.
\newblock In \bibinfo{editor}{Wagner, D.} (ed.)
  \emph{\bibinfo{booktitle}{Advances in Cryptology -- CRYPTO 2008}},
  \bibinfo{pages}{572--591} (\bibinfo{publisher}{Springer Berlin Heidelberg},
  \bibinfo{address}{Berlin, Heidelberg}, \bibinfo{year}{2008}).

\bibitem{SemihomomorphicEncryption}
\bibinfo{author}{Bendlin, R.}, \bibinfo{author}{Damg{\aa}rd, I.},
  \bibinfo{author}{Orlandi, C.} \& \bibinfo{author}{Zakarias, S.}
\newblock \bibinfo{title}{Semi-homomorphic encryption and multiparty
  computation}.
\newblock In \bibinfo{editor}{Paterson, K.~G.} (ed.)
  \emph{\bibinfo{booktitle}{Advances in Cryptology -- EUROCRYPT 2011}},
  \bibinfo{pages}{169--188} (\bibinfo{publisher}{Springer Berlin Heidelberg},
  \bibinfo{address}{Berlin, Heidelberg}, \bibinfo{year}{2011}).

\bibitem{Yao86}
\bibinfo{author}{Yao, A. C.-C.}
\newblock \bibinfo{title}{How to generate and exchange secrets}.
\newblock In \emph{\bibinfo{booktitle}{Proceedings of the 27th Annual Symposium
  on Foundations of Computer Science}}, SFCS '86, \bibinfo{pages}{162--167}
  (\bibinfo{year}{1986}).

\bibitem{Impagliazzo89}
\bibinfo{author}{Impagliazzo, R.} \& \bibinfo{author}{Rudich, S.}
\newblock \bibinfo{title}{Limits on the provable consequences of one-way
  permutations}.
\newblock In \emph{\bibinfo{booktitle}{Proceedings of the Twenty-first Annual
  ACM Symposium on Theory of Computing}}, STOC '89, \bibinfo{pages}{44--61}
  (\bibinfo{publisher}{ACM}, \bibinfo{address}{New York, NY, USA},
  \bibinfo{year}{1989}).

\bibitem{Laudenbach2017}
\bibinfo{author}{Laudenbach, F.}, \bibinfo{author}{Kalista, S.},
  \bibinfo{author}{Hentschel, M.}, \bibinfo{author}{Walther, P.} \&
  \bibinfo{author}{H{\"u}bel, H.}
\newblock \bibinfo{title}{A novel single-crystal \& single-pass source for
  polarisation-and colour-entangled photon pairs}.
\newblock \emph{\bibinfo{journal}{Scientific reports}}
  \textbf{\bibinfo{volume}{7}}, \bibinfo{pages}{7235} (\bibinfo{year}{2017}).

\bibitem{guan2018}
\bibinfo{author}{Guan, J.-Y.} \emph{et~al.}
\newblock \bibinfo{title}{Experimental preparation and verification of quantum
  money}.
\newblock \emph{\bibinfo{journal}{Physical Review A}}
  \textbf{\bibinfo{volume}{97}}, \bibinfo{pages}{032338}
  (\bibinfo{year}{2018}).

\bibitem{bozzio2018}
\bibinfo{author}{Bozzio, M.} \emph{et~al.}
\newblock \bibinfo{title}{Experimental investigation of practical unforgeable
  quantum money}.
\newblock \emph{\bibinfo{journal}{npj Quantum Information}}
  \textbf{\bibinfo{volume}{4}}, \bibinfo{pages}{1--8} (\bibinfo{year}{2018}).

\bibitem{erven2014}
\bibinfo{author}{Erven, C.} \emph{et~al.}
\newblock \bibinfo{title}{An experimental implementation of oblivious transfer
  in the noisy storage model}.
\newblock \emph{\bibinfo{journal}{Nature communications}}
  \textbf{\bibinfo{volume}{5}}, \bibinfo{pages}{1--11} (\bibinfo{year}{2014}).

\end{thebibliography}


\begin{thebibliography}{1}
\expandafter\ifx\csname url\endcsname\relax
  \def\url#1{\texttt{#1}}\fi
\expandafter\ifx\csname urlprefix\endcsname\relax\def\urlprefix{URL }\fi
\providecommand{\bibinfo}[2]{#2}
\providecommand{\eprint}[2][]{\url{#2}}

\bibitem{otps}
\bibinfo{author}{Roehsner, M.-C.}, \bibinfo{author}{Kettlewell, J.~A.},
  \bibinfo{author}{Batalh{\~a}o, T.~B.}, \bibinfo{author}{Fitzsimons, J.~F.} \&
  \bibinfo{author}{Walther, P.}
\newblock \bibinfo{title}{Quantum advantage for probabilistic one-time
  programs}.
\newblock \emph{\bibinfo{journal}{Nature communications}}
  \textbf{\bibinfo{volume}{9}}, \bibinfo{pages}{1--8} (\bibinfo{year}{2018}).

\bibitem{blinov2004}
\bibinfo{author}{Blinov, B.~B.}, \bibinfo{author}{Moehring, D.~L.},
  \bibinfo{author}{Duan, L.-M.} \& \bibinfo{author}{Monroe, C.}
\newblock \bibinfo{title}{Observation of entanglement between a single trapped
  atom and a single photon}.
\newblock \emph{\bibinfo{journal}{Nature}} \textbf{\bibinfo{volume}{428}},
  \bibinfo{pages}{153--157} (\bibinfo{year}{2004}).

\bibitem{Clauser1969}
\bibinfo{author}{Clauser, J.~F.}, \bibinfo{author}{Horne, M.~A.},
  \bibinfo{author}{Shimony, A.} \& \bibinfo{author}{Holt, R.~A.}
\newblock \bibinfo{title}{Proposed experiment to test local hidden-variable
  theories}.
\newblock \emph{\bibinfo{journal}{Phys. Rev. Lett.}}
  \textbf{\bibinfo{volume}{23}}, \bibinfo{pages}{880--884}
  (\bibinfo{year}{1969}).

\end{thebibliography}
\putbib[OTP]
\end{bibunit}

\begin{bibunit}
\section*{Acknowledgements}  
We thank Robert Peterson, Teodor Strömberg, Joshua A. Slater, Fabian Laudenbach, Stefan Zeppetzauer and Maxime Jacquet for discussions and Irati Alonso Calafell and Lee Rozema operating the superconducting detector units.
M.-C.R. acknowledges support from the the uni:docs fellowship program of the University of Vienna. P.W. acknowledges support from the research platform TURIS, the Austrian Science Fund (FWF) through BeyondC (F7113-N48) and NaMuG (P30067-N36), through the European Commission via UNIQORN (no. 820474) and HiPhoP (no. 731473), the United States Air Force Office of Scientific Research via QAT4SECOMP (FA2386-17-1-4011) and Red Bull GmbH. J.A.K. and J.F.F acknowledge support from the Singapore National Research Foundation under NRF Award No. NRF-NRFF2013-01.

\section*{Author Information} Correspondence and requests for materials should be addressed to Marie-Christine Roehsner (marie-christine.roehsner$@$univie.ac.at) and Philip Walther (philip.walther$@$univie.ac.at).

\section*{Appendix}

\subsection{Theory}
\subsubsection{$\mathcal{G}_k$ gate-OTPs}
The presented protocol for $\mathcal{G}_1$ gates may be implemented as a subroutine to realize all possible $\mathcal{G}_k$ gate-OTPs with information theoretic security in a similar fashion to the protocol of \cite{otps}, where subscripts $1$ and $k$ stand for gates with $1$ and $k$ inputs, respectively. All binary inputs to gates are mapped to anti-commuting measurement set $\{M_i\}$,  such that each measurement is composed of separable qubit measurements. Specifically 
 
 \begin{align}
 M_i = \bigotimes_{j=1} ^{2^{k}-1}  \sigma_{ij} ~~ \forall ~ i 
 \end{align}
 
\noindent where $\sigma_{ij} \in \{ \sigma_X, \sigma_Z \}$. Thus all measurements are single qubit operations in one of two bases. Each gate-OTP may be written as 
 
\begin{align}
\rho_G &= \frac{1}{Tr(\mathbb{I})} \left( \mathbb{I} + \frac{1}{\sqrt{2^k}}\sum _{i=1} ^{2^{k}} (-1)^{G(i)} M_i \right) \label{maineq} \\ 
&= \sum_i \frac{1}{2^{k}} \rho_i \\
&= \sum_i \frac{1}{2^{k}} \left( \bigotimes _{j=1} ^{2^{k}-1} \tilde{G_{ij}} \right) 
\end{align}

\noindent where $\rho_i$ is a pure state formed from a tensor product of single qubit states $\tilde{G_{ij}}$. Remarkably, each $\tilde{G_{ij}}$ is a $\mathcal{G}_1$ gate-OTP \cite{otps} and via randomly selecting from the set of possible pure states, the state received by the client is equivalent to the mixed state $\rho_G$ under all measurements. It is thus possible to implement a $\mathcal{G}_k$ gate-OTP using only $\mathcal{G}_1$ states. The probability of correctness $P_k$ of such noisy logic gates is for all inputs

\begin{align}
P_k = \frac{1}{2^{(1+k/2)}}+\frac{1}{2}.
\label{success_prob}\end{align}

 The shared table records random implementations of $\mathcal{G}_1$ gates with measurements in both the $\sigma_X$ and $\sigma_Z$ basis. The protocol presented in the main text allows secure evaluation of the measurement of such states, and thus repeated applications may be used to construct measurement outcomes of $\mathcal{G}_k$ gate-OTPs. The evaluation of all such gates-OTPs may be performed concurrently as the corresponding measurements are separable. We therefore expect the implementation of any $\mathcal{G}_k$ gate-OTP to be completed within an average of $\log _2 \left( 2^k-1 \right) $ rounds of classical communication. A dishonest client, who has not made measurements and instead retained states in a quantum memory, will be in possession of exactly the quantum state intended and described by equation \ref{maineq}, the security of which has been previously shown \cite{otps}. Thus, the delaying of measurements does not allow Bob to obtain additional information regarding the one-time program.

\subsection{Experimental Implementation}
\subsubsection{Source}
A $515$nm cw-laser (Roithner RLTMGL-515-500-2) with a spectral bandwidth of $0.057$nm and a power of $40$mW is used to pump a $30$mm periodically-poled KTP ($KTiOPO_4$) crystal with a poling period of $33.53\mu$m. This is phase-matched for two SPDC processes, one emitting $\ket{H}_s\ket{V}_l$  as well as $\ket{V}_s\ket{H}_l$ with $\lambda_s=785$nm and $\lambda_l=1498$nm. The two down-conversion processes have a different spectral width, thus we  use a narrow bandwidth filter ($0.45$nm) for the photons in the long-wavelength arm. It turns out that it is not necessary to filter the photons in the short-wavelength arm as only photons in the desired wavelength interval will cause coincidences. We found a coincidence rate between Alice and Bob of $10$kHz using a coincidence time window of $6$ns.
Transmission losses between Alice's and Bob's laboratory were measured to be \si{\num{13}\pm \num{2}\percent}.
Using the measured double clicks at one side as well as the detector efficiency and transmission losses we can estimate the percentage of times where more than one photon was emitted finding a value of $0.097\%$. Assuming Bob could use all of these events to improve his probability of success when cheating (i.e. for this percentage of events he has the honest probability of success even when signing two lines) this raises his overall probability of success for a signature run from $0.107\%$ to $0.112\%$. 

\subsubsection{Bell State}
\begin{figure}[t]
\centering
\includegraphics[width=0.45\textwidth]{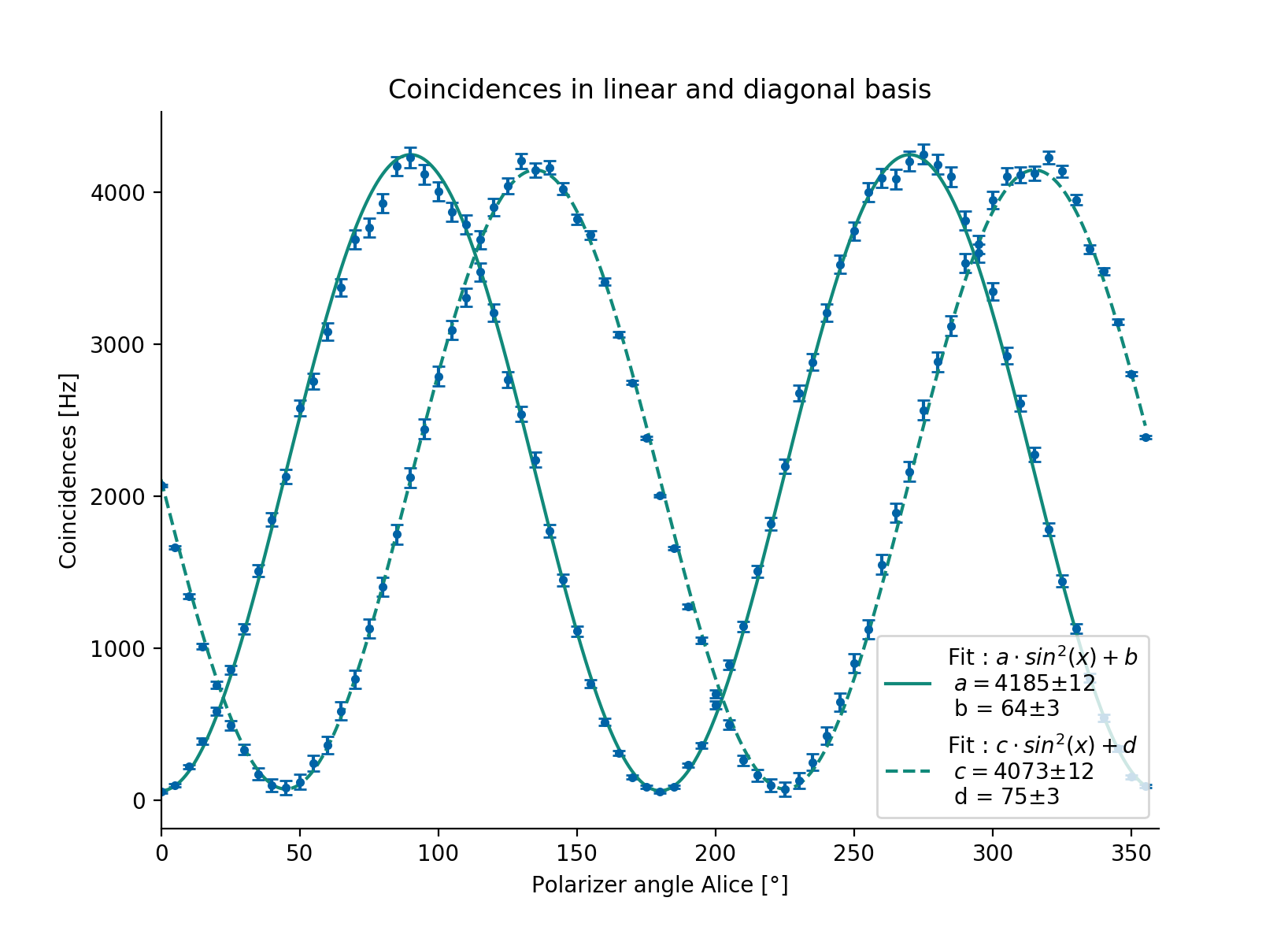}
\caption{Coincidences in linear and diagonal basis: Coincidences in dependence of the polarizer angle in Alice's (he short-wavelength) arm while a polarizer in the long-wavelength arm is fixed at $\ket{H}$ (solid line) or $\ket{+}$ (dashed line). The Visibility is calculated from the sinusodial fit to be $0.974 \pm 0.002$ in the linear basis and $0.965  \pm 0.002$ in the diagonal basis. Blue dots represent experimental data, error bars show one standard deviation and are derived assuming a poissonian distribution. }   
\label{HV}
\end{figure}

\begin{figure}[b]
\centering
\includegraphics[width=0.45\textwidth]{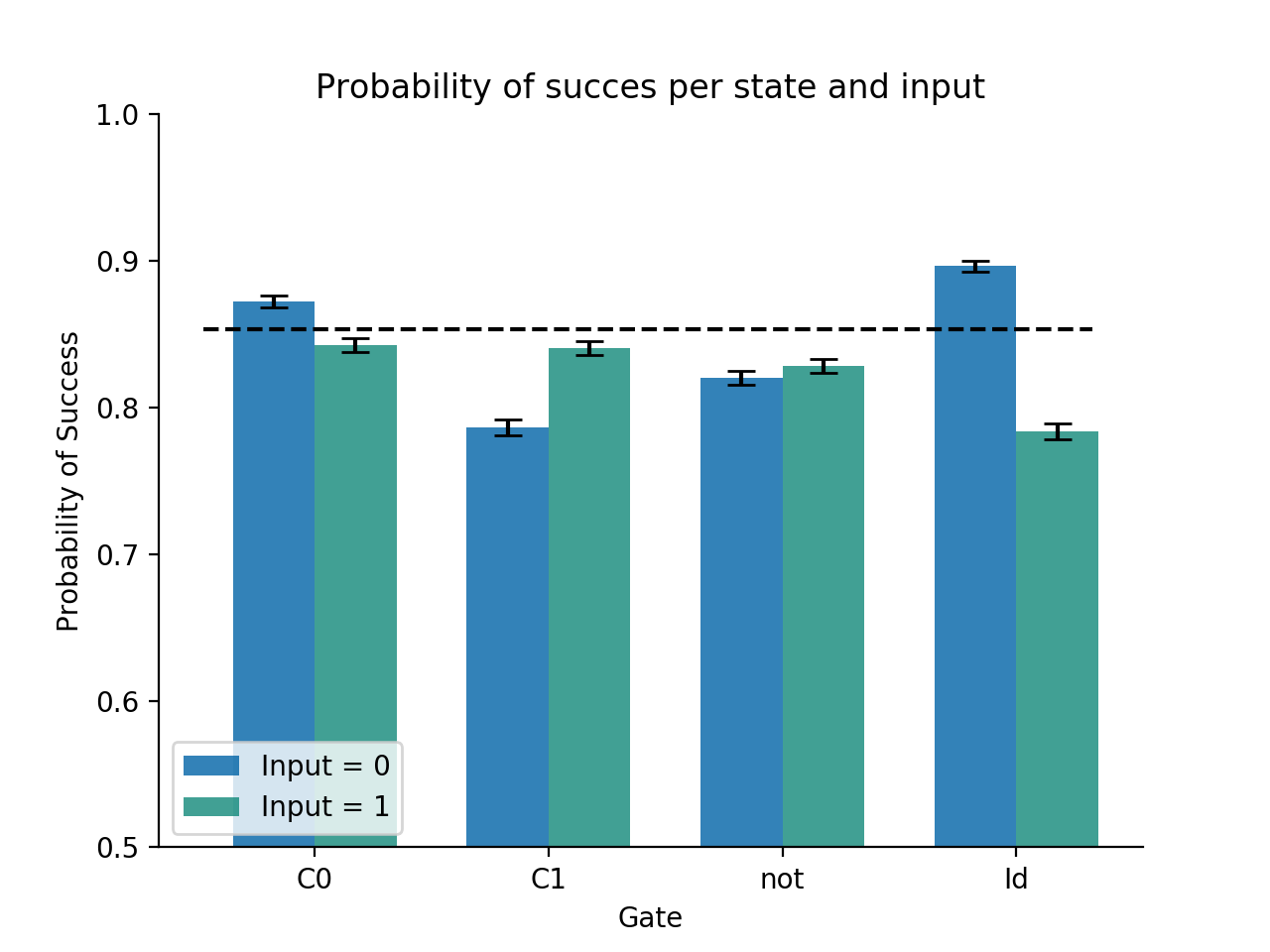}
\caption{Probability of success by state and input. The black dashed line shows the theoretical prediction. The probability of succeess was calculated for all four gates and two inputs each using a sample of $50 000$ lines of the shared table. Error bars show one standard deviation. }                                 
\label{BarPlot}
\end{figure}

Following \cite{blinov2004} we calculated a lower bound on the fidelity $F = \bra{\Psi^-} \rho_{exp} \ket{\Psi^-} $ of the quantum state produced in Alice's lab. We find a value of $F\geq 0.966 \pm 0.003$.
Furthermore in \fig{HV} we show the coincidence counts (measured locally in Alice's laboratory) with respect to the relative angle of polarizes inserted into both arms of the source together with a sinusoidal fit using non-linear least squares.  \fig{BarPlot} shows the probability of success for all four one-bit gates and both possible inputs. 

\subsubsection{Bell inequality violation}
Alice and Bob choose $5000$ lines from their shared table to violate a CHSH-Bell inequality \cite{Clauser1969} as a measure to detect a potential eavesdropper performing a man-in-the-middle attack. They find a Bell parameter of $S= 2.701 \pm 0.042$
violating the classical bound of $S=2$ by  $16.8$  standard deviations.

\subsubsection{Synchronisation}
Alice and Bob use each an identical Time Tagging module (Roithner TTM 8000) to record their detection events. After a measurement run they exchange the timing information on their detected photons to find coincidences. As the clock frequency of the two devices is not identical and the exact starting time of the measurement might vary in the two laboratories they need to find the offset between their respective time stamps. This is achieved using  the switch in the path of the pump laser, before the crystal. Whenever a measurement is started Alice and Bob start their data acquisition with a closed switch. The same signal that triggers the start of the measurement will also trigger the switch to open. Alice and Bob will detect the sharp increase in single photon detections and use this to find a first estimate for the offset in their data. Furthermore, the first $100$ ms  of each measurement run are used by Alice and Bob to find the correct offset and calibrate their timing offsets. After this period the switch is closed and opened again, indicating the start of the distribution of a shared table. 
\putbib[OTP1]
\end{bibunit}

\end{document}